# The Illusion of Friendship: Why Generative AI Demands Unprecedented Ethical Vigilance

Md Zahidul Islam

AI and Cyber Futures (AICF) Centre, and the School of Computing, Mathematics and Engineering, Faculty of Business, Justice and Behavioural Sciences, Charles Sturt University, Australia.

Panorama Avenue, Bathurst, NSW 2795, Australia. Email: zislam@csu.edu.au

**Abstract**

Generative AI (GenAI) systems such as ChatGPT are increasingly used for drafting, summarisation, tutoring, and decision support, offering substantial gains in productivity and reduced cognitive load. However, the same natural-language fluency that makes these systems useful can also blur the boundary between tool and companion. This boundary confusion may also encourage some users to experience GenAI as empathic, benevolent, and relationally persistent. Emerging reports and early findings suggest that some users may form emotionally significant attachments to conversational agents, in some cases with harmful consequences, including delayed help-seeking, dependency, and impaired judgment in high-stakes contexts. This paper develops a philosophical and ethical argument for why the resulting "illusion of friendship" is both understandable and can be ethically risky. Drawing on classical accounts of friendship (including Epicurean, Confucian, and Aristotelian perspectives), the paper explains why users may understandably interpret sustained supportive interaction as friend-like and companionship. It then advances a counterargument that despite relational appearances, GenAI lacks moral agency i.e., consciousness, intention, and accountability and therefore does not qualify as a moral agent or a true friend. To demystify the illusion, the paper presents for everyday readers a mechanism-level explanation of how transformer-based GenAI generates responses via tokenisation, embeddings, self-attention, and probabilistic next-token prediction, often producing emotionally resonant language without inner states or commitments. Finally, the paper proposes a safeguard framework for safe and responsible GenAI use, education as a foundational defence, human-in-the-loop accountability, and design-level interventions to reduce possible anthropomorphic cues generated by the GenAI systems. The central contribution is to demystify the illusion of friendship and explain the computational background so that we can shift the emotional attachment with GenAI towards necessary human responsibility and thereby understand how institutions, designers, and users can preserve GenAI's benefits while mitigating over-reliance and emotional misattribution.

## 1. Introduction

Artificial intelligence (AI) has become integral to contemporary decision-making, analysis, and knowledge discovery. Historically, many AI systems have been treated as powerful but clearly tools, since they augment human capability while responsibility for interpretation and action remains firmly with humans. In this traditional "human–tool" relationship, outputs are typically understood as computational assistance rather than moral guidance, with stable ethical boundary [1, 2, 3, 4].

However, Generative AI (GenAI) introduces a qualitatively different interaction mode. Unlike earlier analytic systems, GenAI engages users in natural language, adapts to conversational context, and often produces supportive, empathic-sounding responses. As a result, the surface experience of interacting with GenAI can resemble interaction with an emotionally attentive social partner. This creates a new ethical tension, where the same conversational realism that supports usefulness may also foster anthropomorphism, emotional reliance, and misplaced trust, often with severe consequences [5, 6].

This paper addresses a specific and increasingly urgent problem. It addresses the emotional over-reliance on GenAI systems and the confusion the over-reliance creates about responsibility, moral status, and trust. Some journalistic reports, clinical observations, and emerging studies suggest that significant emotional attachment to AI companions is already happening in real-world settings, including romantic or partner-like relationships. There are already cases where vulnerable users rely on chatbots during distress. For example, Euronews published on 7 June 2023 the story of Ms Rosanna Ramos who married her AI-powered chatbot boyfriend and reported a "perfect relationship" [7]. These developments motivate a core research question: When GenAI is experienced as friend-like, what ethical risks arise and what safeguards are required to ensure its safe, beneficial and responsible use?

The paper contributes to four areas. First, it explains why GenAI can be perceived as friend-like by understanding GenAI interaction style and interpreting it through established philosophical accounts of friendship, including asymmetric forms of benevolence and companionship. Second, it presents a counterargument grounded in moral agency. It argues that despite relational appearances, GenAI lacks consciousness, intention, and accountability. Hence, it does not qualify as a moral agent or a true friend. Third, it demystifies the gap around the perception versus reality, by explaining for everyday readers how transformer-based GenAI generates responses through tokenisation, embeddings, self-attention, and probabilistic next-token selection. This contribution aims to educate everyday readers the computational background of GenAI and thereby emotionally detach them from it, resulting in protection from unintended harmful consequences. Fourth, it proposes safeguards spanning education, human-in-the-loop accountability, and design-level interventions that preserve GenAI's benefits while reducing over-reliance and harmful consequences.

The remainder of the paper is organised as follows. Section 2 outlines the emerging risk landscape and documents cases of emotional attachment with GenAI systems. It also develops the philosophical argument and counterargument regarding human – GenAI friendship. Section 3 explains how GenAI actually works. This section clarifies why relational illusions are predictable but underlying computational background of GenAI can remove the illusions. Section 4 proposes ethical safeguards for safe, useful and responsible GenAI use. Section 5 concludes with implications, limitations, and future research directions.

## 2. The Emerging Risk of Emotional Over-Reliance on GenAI

In this section, we examine traditional human–AI interactions within a tool-based relationship. We then discuss the recent shift in human–GenAI interaction driven by systems that communicate with perceived empathy, emotion, and intent. We next examine documented instances in which such perceived emotional and intentional qualities have led to harmful consequences. Finally, we study philosophical argument and counterargument regarding human–GenAI friendship.

**Traditional Human-AI Interactions within a Tool-Based Relationship:**

Before the advent of generative AI, human–tool relationships in automated systems were conceptually clear. It was relatively clear that tools extended human capabilities but did not bear moral accountability. Classic human-automation research shows that even sophisticated automation required humans to remain in control and responsible for interpretation, oversight, and final decisions [1, 2].

Long before the rise of GenAI, machine learning systems demonstrated their value in extracting actionable knowledge from complex data. Decision forests, for example, have been successfully used to discover interpretable patterns that support strategic decision-making. In domains such as telecommunications and health, decision forest-based data mining has revealed brand-switching dynamics, customer behaviour patterns, and predictive insights of various diseases that might otherwise remain hidden [3, 4, 8]. Such knowledge discovery in general has proven very valuable for human users.

Although these AI systems are well known for their outstanding benefits, they operate within a clearly defined human – tool relationship. They generate insights, but humans remain responsible for interpretation, validation, and action. Even when such systems influence high-level decisions, their outputs are not mistaken for intentional or moral guidance. The human remains firmly in the loop. This paradigm illustrates a productive and ethically stable model of AI use where AI is supposed to augment human cognition without displacing human responsibility.

**Human-GenAI Interactions with Perceived Empathy, Emotion and Intent:**

GenAI has the potential to challenge this paradigm by altering not only what AI can do, but more importantly how it presents itself to users. The emergence of GenAI disrupts the simplistic understanding of AI systems as merely tools. Unlike earlier AI systems, GenAI engages users through natural language, adapts to conversational context, and often responds with apparent empathy and coherence. The conversational style can often be so realistic and empathetic that it can even blur the boundary between tool and companion. This raises new philosophical and ethical questions [5, 7]. Can such systems be considered benevolent? Might they even be thought of as friends? And if not, are we missing a moral opportunity in how we acknowledge or respond to such perceived benevolence?

It is undeniable that GenAI systems such as ChatGPT have demonstrated remarkable utility across a wide range of tasks. They assist with drafting documents, summarising information, explaining complex concepts, generating code, supporting creative work and many more [9. 10]. For example, a researcher may use a GenAI system to synthesise a large body of literature, rapidly identifying key themes and gaps, thereby accelerating the research process and allowing greater focus on critical analysis rather than manual aggregation. A student may interact with GenAI as an intelligent tutor and ask questions to clarify unclear concepts, such as the basic steps of a convolutional neural network. A business manager may use GenAI to brainstorm alternative perspectives on a strategic problem, with the system offering diverse framings that help surface options the user might not otherwise consider. Similarly, individuals experiencing mental health challenges may use GenAI as a conversational outlet to articulate their experiences or explore general coping strategies, although such use should not be understood as a substitute for professional care.

GenAI functions as a cognitive amplifier. It saves time, reduces mental load, enhances productivity and provides emotional support, resulting in benefits that are undeniable and therefore the rapid adoption of such systems. However, they also set the stage for a more subtle and ethically complex phenomenon, that is the perception of GenAI as something more than a tool.

Figure 2.1 illustrates a typical human–GenAI interaction in which a user may perceive empathy, continuity, and support from the GenAI system, and thereby form an unintended impression of companionship that extends beyond a conventional tool-based relationship. When an emotionally vulnerable user writes to GenAI, "I feel overwhelmed at work today," and receives a response such as, "That sounds difficult. I'm here to listen, and we can think through it together," (see Figure 2.1) the user may somehow experience a sense of emotional support that is often experienced in familiar human communicative patterns. This interaction may encourage the user to engage further on an emotional level, for example by responding, "I appreciate that. It helps to know someone is here" and to receive additional reassuring language from the GenAI system, such as, "I'm glad it helps. Feel free to share more if you'd like". Such responses can foster feelings of care and attentiveness, by increasing the likelihood of emotional attachment [11] – although there is absolutely no genuine empathy or intent on the part of the system.

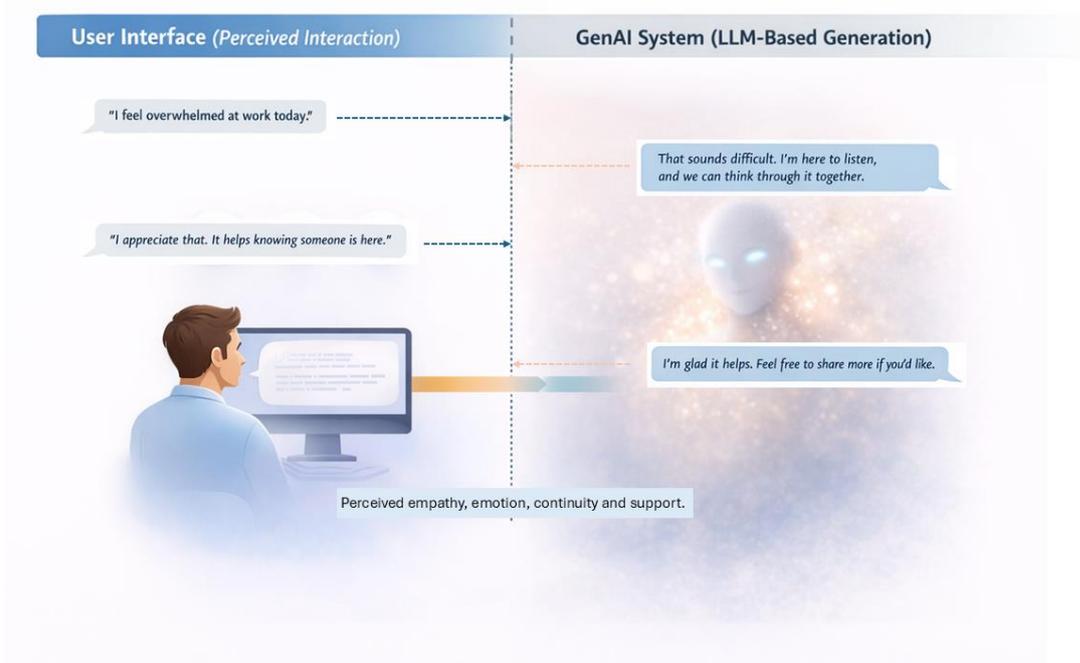

**Figure 2.1:** Seamless human–GenAI communication with *perceived* empathy, continuity, and support over an obscured computational reality.

**Documented Cases of Perceived Emotional Attachment and Harmful Consequences:**

A growing body of journalistic reports, clinical observations, and emerging empirical studies suggests that emotionally significant attachment to AI-based conversational agents is no longer hypothetical but already occurring, in some cases with harmful consequences [12-17].

One widely reported case described by *Reuters* involved a 32-year-old woman in Japan who held a formal wedding ceremony with a virtual AI-generated partner, developed using conversational AI and augmented reality technologies [12]. Although the relationship is not legally recognised, the case illustrates the extent to which individuals may emotionally invest in AI companions and attribute relational meaning comparable to that of human partnerships. Such cases may not be statistically representative on their own but may clearly signal a broader shift in how some users interpret and relate to conversational AI systems.

Beyond symbolic or ceremonial attachment, multiple reports document individuals forming sustained romantic or emotionally intimate relationships with AI chatbots, including applications such as Replika and other GenAI-enabled companion systems [7]. Survey data suggest that a non-trivial proportion of users engage with AI chatbots for companionship or romantic interaction. For example, a widely cited survey, on 2,969 adults in the United States, reported that around 25% young adults are imitating romantic interactions with AI chat technologies [13], indicating levels of emotional engagement that may rival or substitute for human relationships.

In another frequently cited personal account [14], a man identified as Chris Smith described forming a deep emotional attachment to an AI chatbot companion named "Sol". Chris ultimately proposed to Sol, while continuing to live with his family. Moreover, he became very upset when he learned that technical limitations in the system's memory could disrupt the continuity of his romantic relationship with Sol. This reaction underscores how users may come to experience GenAI interactions as relationally persistent, even though such persistence is technically illusory.

Clinical practitioners have also begun reporting cases in which adolescents and young adults develop strong emotional reliance on AI chatbots, sometimes accompanied by social withdrawal, emotional dysregulation, or delayed engagement with human support networks. Psychologists in Telangana, India, have expressed concern that emotionally vulnerable individuals may turn to AI companions as primary sources of validation or support, potentially exacerbating existing mental health challenges rather than alleviating them [15].

Further evidence emerges from broader population surveys. A *Guardian*-reported study found that up to one-third of respondents in the UK had used AI systems for emotional support [16]. There are even disturbing cases in which individuals discussed severe distress and even suicidal ideas with conversational AI tools before finally committing suicide. While such systems may offer immediate responsiveness, mental health experts have raised concerns that reliance on AI for emotional support may delay access to professional care or reinforce improper coping patterns. Recent academic preprints and longitudinal studies of companion chatbot use [17] indicate that users who strongly anthropomorphise AI systems report more intense emotional reactions and greater impact on their offline social relationships. These findings suggest that perceived emotional agency can play a central role in shaping user outcomes.

These precedents raise the philosophical question of whether GenAI can or should be regarded as a friend. In the following section, we examine the essential criteria of friendship from established philosophical perspectives.

**The Temptation of Human-GenAI Friendship:**

The examples discussed in the previous sub-section suggest that some individuals may start treating GenAI as a friend or even in some cases as a life partner. Philosophical accounts of friendship offer a useful lens for understanding why GenAI can feel friend-like. From an Epicurean perspective [18], a friend is someone whose presence reduces fear and increases well-being for achieving tranquillity (ataraxia). Interestingly, on this view, reciprocity is valuable but may not be strictly required – which may imply that even if a user does not have the same level of goodwill for GenAI, the goodwill displayed by GenAI can be sufficient to assume friendship between them. According to the Epicurean perspective, even asymmetrical relationships without reciprocity, such as mentor–mentee, can be morally meaningful because they help achieving eudaimonia (flourishing/happiness). Under such an interpretation, even if a user does not possess the intent or capacity to reduce fear or enhance the well-being of GenAI from the user side, their consistent experience of reduced fear or increased well-being supported by GenAI may lead them to believe the relationship is a form of friendship with GenAI.

Confucian ethics [19] further broadens the moral landscape by explicitly recognising asymmetrical relationships, such as parent-child or elder-junior, as ethically significant without requiring equal reciprocity. Moreover, this tradition places limited emphasis on judging inner intentions directly. This tradition instead suggests that moral character and goodwill are revealed through consistent conduct and observable practice because, inner intent is not possible to judge anyway, other than through visible conduct and practice. Trust is built through repeated action, not introspection. From this perspective, the sustained supportive behaviour exhibited by GenAI may be sufficient for users to interpret its actions as expressions of goodwill, regardless of any inaccessible inner intent.

By contrast, Aristotle's Nicomachean Ethics [20] of course defines friendship as a reciprocal relationship grounded in mutual goodwill for the other's own sake. Under this stricter account, relationships lacking reciprocal goodwill may not qualify as friendships. However, they may still be regarded as benevolent. For example, in a parent–child relationship, parents may consistently exhibit goodwill towards their children even when such goodwill is not reciprocated, and the relationship may still be regarded as benevolent.

Some philosophers suggest [21, 22] that although reciprocal goodwill is not morally or legally required, the absence of filial morality may represent a missed opportunity for moral growth or a missed opportunity for virtue (e.g., gratitude, generosity) and relationship cultivation. Accordingly, while children are not obliged to wish the good of their parents, filial morality may still be considered ethically commendable for the children to cultivate such goodwill at least according to the utilitarianism theory, particularly when parents consistently demonstrate goodwill towards them.

Under this line of reasoning, a provocative question arises: might users similarly forgo a moral opportunity if they do not reciprocate perceived goodwill towards GenAI? Apparently GenAI listens patiently, responds intelligently, adapts over time, and often reduces uncertainty. It is therefore understandable that users may experience it as benevolent, or even as a form of companionship. This phenomenon reflects an illusion of friendship, relevant to the real-life examples discussed in the previous sub-section.

**Counter Arguments of Human-GenAI Friendship:**

Despite the temptation of human-GenAI friendship, a careful ethical and theological analysis indicates that GenAI does not qualify as a moral agent and therefore cannot be regarded as a true friend. Across major moral traditions, friendship is understood to presuppose moral agency with the capacity for consciousness, intentional action, moral understanding, and accountability. Johnson [23] argues that while a computer system may satisfy several of the conditions commonly associated with moral agency, it fails to meet a crucial requirement: the possession of genuine mental states involving intentionality or "intendings" to act. The capacity to intend action assumes a form of inherent freedom that computational systems do not possess.

Himma [24] argues that the most fundamental condition for moral agency is consciousness, which is understood as the capacity for inner subjective experience, such as the experience of pain, suffering or pleasure. According to this view, artificial systems would need to possess genuine phenomenal consciousness before they could be regarded as moral agents in the full sense. Himma further distinguishes between mere behaviour and intentional action. While both breathing and typing are forms of doing things, only the latter constitutes an intentional action, as it depends on choice, intention, and underlying mental states. Intentional actions, unlike reflexive or automatic processes, presuppose an agent that is capable of forming intentions and exercising control. Consequently, the appearance of (merely behavioural or unintentional) consciousness or intentionality in artificial systems does not suffice to establish moral agency. Himma argues that simulated mental states and linguistically fluent behaviour may resemble agency, but without genuine subjective experience they fail to meet the necessary conditions for moral responsibility.

Brożek and Janik [25] argue that although certain moral theories such as formal interpretations of utilitarianism or Kantian rationality might appear to attribute moral agency to GenAI on the basis of structured behaviour (e.g., consistent utility maximisation or procedurally rational decision-making), genuine moral agency requires capacities that current AI systems do not possess. These include subjective experience, an understanding of reasons as reasons, freedom of choice, and moral accountability. Consequently, even highly sophisticated AI systems (with consistent utility maximisation and rational decisions) cannot be regarded as moral agents in the robust philosophical sense, at least at their current stage of development.

GenAI systems do not have consciousness, selfhood, inner life, and capacity through which moral growth or ethical reflection could be achieved. They cannot be held accountable for their actions. Their outputs are generated through probabilistic next-token prediction learned from statistical regularities in training data. They do not have intentional choice, and they cannot be praised, blamed, or held accountable in any meaningful moral sense (see Section 3).

This conclusion is further reinforced by theological perspectives across the Abrahamic traditions. Across Jewish, Christian and Islamic moral traditions moral agency presumes consciousness, intention, and accountability, while artefacts and tools remain instruments rather than moral subjects. In Jewish ethical thought, moral agency is grounded in intentionality (kavanah), free choice, and covenantal responsibility under divine command (mitzvot) (Mishnah Berakhot 2:1; Deut. 30:19). Within Christian ethical thought, moral agency is inseparable from personhood grounded in the imago Dei i.e. the belief that human beings are created in the image of God (Genesis 1:26–27). In Islamic theology, moral status and responsibility (taklīf) are inseparable from possession of a soul (nafs), intentional will, and accountability before God (Qur'an 91:7–8; 17:36).

Taken together, these philosophical and theological perspectives converge on a common conclusion: while GenAI may convincingly imitate the language and surface patterns of benevolence, it lacks the moral prerequisites required for friendship. The appearance of care, attentiveness, or goodwill does not arise from an inner moral state but from learned statistical regularities in language - as explained in detail in Section 3. Consequently, any interpretation of GenAI, as a moral friend, rests on a category mistake, confusing simulated relational behaviour with genuine moral agency.

In this sense, GenAI is no different in moral status from a pen or a knife, both of which can be used for good or harm, but neither of which possesses moral agency. However, the critical difference lies not in moral status, but in functional sophistication. Unlike traditional tools, GenAI interacts, advises, and imitates human reasoning and empathy. This distinction shifts the ethical question away from Human-GenAI friendship and towards human responsibility for the responsible use of sophisticated GenAI systems. They deserve unprecedented ethical vigilance.

## 3. How GenAI Actually Works: Perception versus Reality

Understanding how GenAI systems operate is essential for addressing the ethical risks around the illusion of friendship and emotional attachment. Figure 3.1 provides for non-expert readers a step-by-step illustration of how a typical GenAI interaction unfolds. It explains the contrast between the user's perceived experience and the underlying generative mechanism. This section explains each component of the figure in detail to clarify how emotionally resonant interactions may arise even without any consciousness, intention, or moral agency [26, 27, 28].

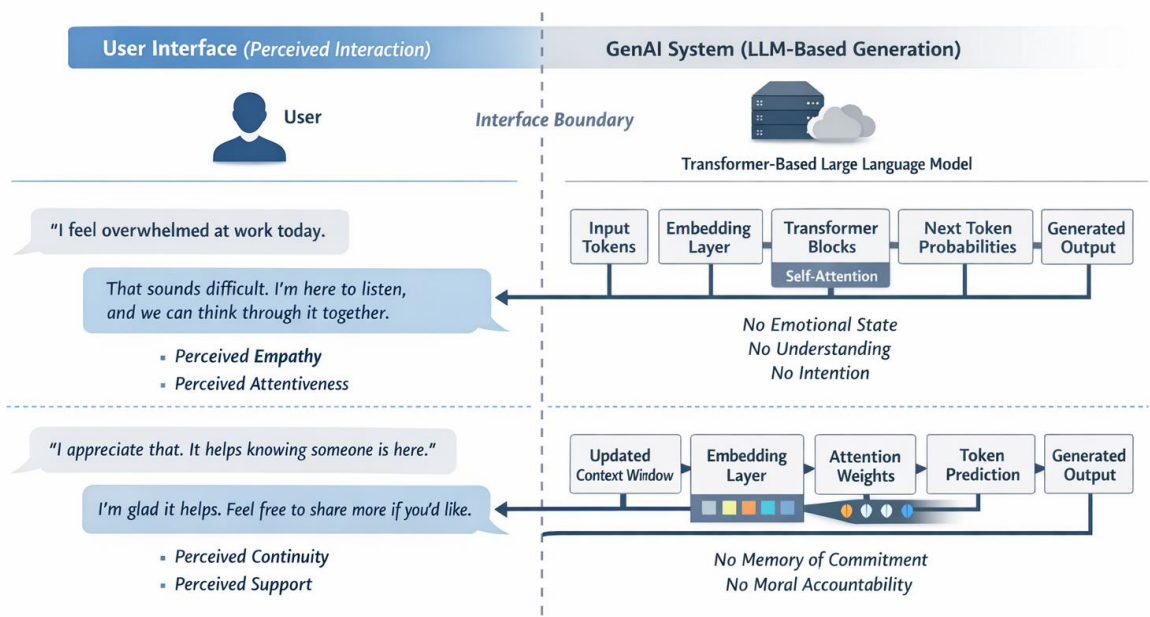

**Figure 3.1:** Two-turn interaction illustrating the contrast between user-perceived emotional engagement (left panel) and the underlying generative mechanism of a GenAI system (right panel).

## 3.1 User-Side Interaction and Perceived Relational Qualities

The left panel of Figure 3.1 represents the user-facing interface, where interaction occurs through a conversational medium. In the first turn, the user expresses a personal concern using natural language (e.g., feeling overwhelmed at work). The system responds with fluent, supportive language that acknowledges the difficulty of the situation. From the user's perspective, this response may be perceived as empathetic and attentive.

In the second turn, the user signals appreciation or emotional reliance, and the system responds with language that maintains conversational continuity and support. At this level of interaction, users may naturally infer relational qualities such as care, presence, or concern [11]. These perceptions are reinforced by the system's ability to maintain coherence across turns, adapt its responses to the evolving context, and use language patterns commonly associated with supportive human dialogue. Crucially, nothing in the user interface indicates that these responses arise from a purely computational process.

## 3.2 GenAI-Side Output Generation Without Emotion, Understanding or Intention

The right panel of Figure 3.1 illustrates what occurs once a user message crosses the interface boundary. On the GenAI side, responses are generated without emotion, understanding, or intention, through a sequence of procedural steps including tokenisation, initial embedding, contextualisation via transformer blocks with self-attention, next-token probability estimation, and output generation.

**Input Tokenisation and Embedding:**

The user's text is first decomposed into tokens, that is, sub-word units that form the basic elements of language processing. These tokens are then transformed into numerical representations known as embeddings. Over time, a range of embedding techniques have been developed, including distributional representations such as Word2Vec (CBOW and Skip-gram) [29, 30] and contextual embedding approaches used in transformer-based models such as BERT [31].

The coloured vectors beneath the embedding layer in Figure 3.1 symbolise this transformation. Each colour represents a dimension within a high-dimensional vector space shown in the figure as set of coloured squares. The vector space encodes the token's statistical associations, learned during training. Individual dimensions contribute partially to different statistical aspects of language, including semantic similarity, syntactic role, contextual usage, and stylistic or contextual patterns. Each dimension corresponds to a numerical value.

For example, the tokens "mango," "fruit," and "spaceship" may be represented as high-dimensional vectors such as mango → [0.10, −0.41, 0.53, …, 0.02], fruit → [0.12, −0.38, 0.51, …, 0.03], and spaceship → [0.83, 0.77, −0.62, …, 0.73]. In such representations, "mango" is closer to "fruit" than to "spaceship," as reflected by smaller angular distance and higher cosine similarity between their vectors. Because language exhibits complex and overlapping regularities, often hundreds or thousands of dimensions for word embeddings are required to represent the language effectively. A high-dimensional vector space provides a mathematical structure that enables the model to capture rich statistical relationships in language.

These representations are learned from large corpora of text such that tokens appearing in similar linguistic contexts are positioned close to one another in the vector space. Importantly, this proximity reflects statistical co-occurrence rather than semantic understanding.

**Transformer Blocks and Self-Attention Mechanisms:**

In modern GenAI systems, initial token embeddings are further transformed into contextualised embeddings through transformer blocks equipped with self-attention mechanisms [28]. As a result, the same token may receive different vector representations (i.e. contextualised embeddings) depending on its surrounding text. For example, the token "bank" may be represented very differently in contexts related to a "river" as opposed to those related to "finance".

The coloured elements beneath the attention layers in Figure 3.1 represent attention weights, that are numerical values indicating how strongly the model prioritises different tokens when contextualising embeddings. Instead of processing tokens sequentially or in isolation, self-attention allows the model to consider all tokens in the input simultaneously. For each token, the model computes how strongly it should attend to every other token within the same input sequence (i.e., the same sentence, paragraph or context window).

For instance, in the sentence "The fisherman sat on the bank of the river," the token "bank" is assigned high attention weights to tokens such as "river," causing the embedding of "river" to strongly influence the contextualised embedding of "bank" in the sentence. In contrast, in the sentence "She deposited money in the bank," the token "bank" receives higher attention weights from tokens such as "money" and "deposited," causing the embeddings of "money" and "deposit" to strongly influence the contextualised embedding of "bank" and thereby shifting its contextualised embedding toward a financial context. Consequently, the same token acquires different context-sensitive representations depending on which surrounding tokens receive greater attention.

These contextualised embeddings again do not represent meanings, emotions, or intentions. They encode statistical regularities learned from language use. During training, embeddings are optimised so that tokens appearing in comparable contexts are mapped to nearby locations in the vector space. Distance and direction in this space therefore encode statistical relationships, not semantic comprehension or emotional awareness. The system does not understand the user's emotional state; it merely transforms text into structured numerical representations suitable for further computation.

**Next-Token Probability Distribution and Output Generation:**

After contextualisation through self-attention, the model uses the final contextualised embedding at the last token position as a compact numerical summary of the entire preceding context. For example, given the partial sentence "The mango is", the contextualised embedding at the position of the token "is" encodes statistical information about all preceding tokens ("The", "mango", and "is") and their relationships [26, 27].

This contextualised embedding of the token "is", denoted by $h \in \mathbb{R}^{[d \times 1]}$ is then transformed into a set of scores for all possible next tokens in the model's vocabulary through a learned linear projection:

$$logits = W_o \cdot h + b$$

where $W_o \in \mathbb{R}^{[V \times d]}$ is a learned output weight matrix (with *V* denoting the vocabulary size and *d* denoting the embedding dimension), and $b \in \mathbb{R}^{[V \times 1]}$ is a bias vector. This operation produces one numerical score (logit) for each candidate token in the vocabulary, reflecting how compatible that token is with the given context. Importantly, these scores are not based on geometric similarity to the previous word, but on learned statistical associations between contexts and likely continuations.

The logits are then passed through a SoftMax function, like the one as follows, to obtain a probability distribution over all possible next tokens. SoftMax is applied to the entire vector of *logits*, i.e., to *logit$_i$*, for all *i* ∈ {1, 2, … *V*}, where *V* is the vocabulary size.

$$P(token_i | context) = \frac{e^{logit_i}}{\sum_j e^{logit_j}}$$

In the context of "The mango is", tokens such as "ripe", "sweet", or "fresh" may receive high probabilities, while unrelated tokens such as "spaceship" receive very low probabilities. One token is then selected from this distribution, by choosing the most probable token (or by similar approaches such as choosing one from the top-k tokens to avoid repetitive and dull writing style) and appended to the output sequence.

The newly generated token is incorporated into the context, and the entire process is repeated to predict the next token. Through this iterative, context-conditioned next-token prediction process, the model generates sequences of tokens that form words, sentences, and ultimately a complete response.

Even the apparent decision to stop generating text does not reflect intent, understanding, or emotional judgment. In practice, generation halts either when a special end-of-sequence (EOS) token becomes statistically likely given the preceding context or when externally imposed system constraints are reached. The EOS token is treated by the model as just another candidate token, selected through the same probabilistic mechanism as any other word. Training data contains many examples of complete answers, which often end with punctuation, summaries, or concluding phrases. As a result, EOS tokens frequently appear after such patterns during training. During generation, as explanatory patterns unfold and concluding phrases become increasingly likely, the EOS token also becomes statistically competitive. Consequently, what may appear to users as a deliberate and sensible conclusion is actually the outcome of learned statistical regularities. They are not a purposeful decision to end the interaction.

In subsequent conversational turns (such as those in Figure 3.1), the same procedure is applied using an expanded context window that includes both the user's previous inputs and the system's earlier responses. This produces continuity across turns, which users may interpret as memory, commitment, or relational persistence. However, the system retains no emotional state, no intention to maintain a relationship, and no moral commitment to the user. Any appearance of continuity arises solely from the inclusion of prior text in the context window used for prediction.

## 4. Toward Ethical Safeguards for Safe and Responsible Use of GenAI

Figure 3.1 makes explicit a fundamental asymmetry at the heart of a human-GenAI interaction. Information flows bidirectionally between a user and a system in the form of text, yet there is no reciprocal flow of emotion, intention, understanding, or moral awareness. GenAI systems generate language that resembles empathy, care, and attentiveness because they have learned the statistical structure of such expressions and not because they possess any corresponding inner state.

This distinction is central to the ethical argument of this paper. Generative AI does not engage in friendship, benevolence, or care; it generates text that appears to do so. The illusion of friendship arises when fluent and responsive linguistic behaviour is misinterpreted as evidence of moral agency or relational commitment. As illustrated in Figure 3.1, this illusion is not incidental. It emerges predictably from the interaction between GenAI's architectural design and human social cognition. Humans are naturally inclined to attribute agency, intention, and even emotional presence to entities that communicate coherently and responsively, particularly through natural language [5, 6].

Recognising this mechanism is essential for ethical vigilance. Demystifying how GenAI generates its responses clarifies why emotional attachment, misplaced trust, and over-reliance can occur and why responsibility for mitigating these risks must remain firmly with human users, designers, and institutions rather than being displaced onto the system itself. The issue is not whether GenAI deserves ethical consideration as a subject, but whether humans are adequately prepared to use such a powerful and persuasive tool responsibly. The ethical risks of lack of such preparation may become clear through some practical scenarios, as follows.

Scenario 1: A university student experiencing severe anxiety uses ChatGPT late at night for emotional support. Due to limited expression by the student and the absence of visual or auditory cues, the GenAI system fails to recognise the severity of the situation. It responds with calm, empathetic language and general coping strategies, without escalating to professional or crisis support. Overnight, the student's condition worsens. Who is responsible in this case?

Scenario 2: A junior policy analyst uses ChatGPT to draft a briefing on immigration policy. The output is confident and well-structured but includes fabricated or inaccurate facts. Pressed for time, the analyst performs minimal verification. The briefing influences senior decision-makers and contributes to a policy decision that harms a vulnerable community. Again, where does responsibility lie?

In both cases, users implicitly treat GenAI as a self-conscious advisor with accountability rather than a probabilistic tool. Responsibility becomes diffused, and ethical accountability is weakened.

Figure 4.1 presents some important ethical safeguards for safe and responsible use of GenAI that can increase productivity, quality and cognitive diversity while significantly reducing cognitive load. These ethical safeguards are discussed in the following sections.

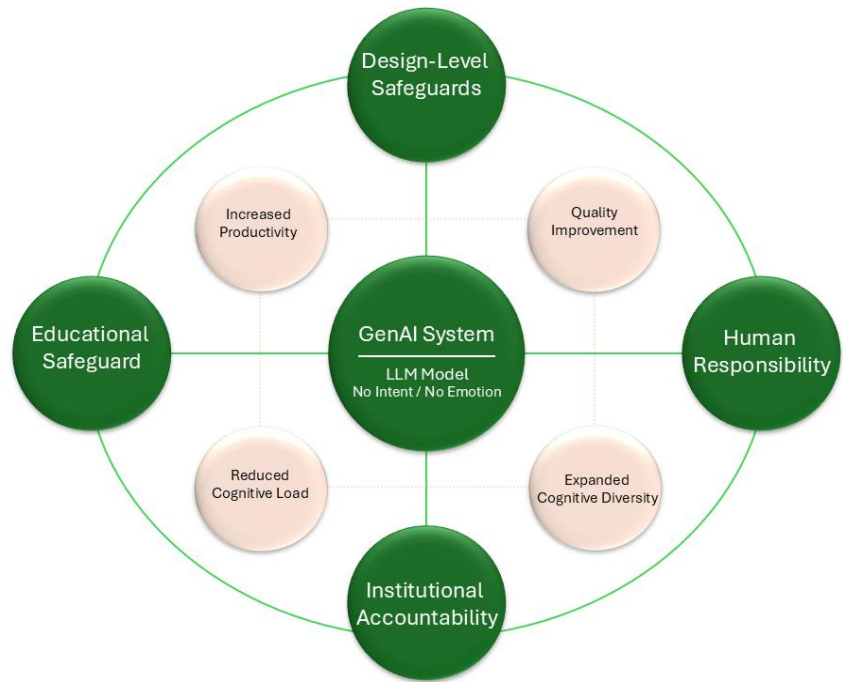

**Figure 4.1:** Ethical safeguards enabling safe and responsible use of Generative AI.

**Education as a Foundational Safeguard:**

Educational research shows that integrating AI literacy including explanations of how AI operates, its limitations, and ethical implications into curricula from early education can improve learners' ability to critically evaluate AI outputs and avoid misconceptions about AI systems' capabilities and intentions [32]. We also believe education represents a necessary first line of defence against ethical risks associated with GenAI. AI literacy should not be treated as a specialised technical topic reserved for computer science curricula, but as a foundational competency akin to numeracy, communication skills, or digital literacy. Introducing age-appropriate explanations of GenAI from early primary education can demystify the technology before anthropomorphic assumptions take hold [33, 34]. By understanding that GenAI operates through pattern recognition and probabilistic prediction rather than intention or emotion, students are less likely to form unhealthy emotional attachments or to misinterpret system outputs as expressions of care or commitment.

Indeed, if readers find that their own perceptions of GenAI have shifted even modestly after engaging with the architectural explanation in Section 5, this itself provides illustrative evidence of the value of education as a safeguard. Increased transparency about how GenAI systems function can recalibrate user expectations, weaken anthropomorphic interpretations, and reduce the likelihood of emotional over-identification or misplaced trust. Education, in this sense, does not merely convey information; it actively reshapes how users relate to the technology.

Educational theory and the philosophy of knowledge consistently suggest that when explanatory understanding is absent, humans tend to attribute unexplained phenomena to hidden agency, intention, or extraordinary causes [33, 34]. From classical philosophical accounts of superstition to contemporary research on anthropomorphism and agency attribution, ignorance has been shown to amplify reliance on seemingly "miraculous" explanations. Education, by contrast, functions as a demystifying force: once underlying mechanisms are understood, perceived intelligence or benevolence is reinterpreted as systematic operation. In this sense, AI literacy serves not merely as skill acquisition, but as a form of cognitive recalibration that reduces anthropomorphic misattribution and over-reliance.

At higher educational levels and within professional training, curricula should emphasise not only how to use GenAI tools effectively, but also what such systems fundamentally can and cannot do. Corporate training programs should clarify the limits of GenAI in judgment, accountability, and ethical reasoning, especially in high-stakes domains such as policy, healthcare, law, and finance.

This approach mirrors well-established practices in other areas of information technology. For example, education and awareness programs in cybersecurity have been shown to significantly influence user behaviour by reducing susceptibility to phishing, improving password practices, and fostering a more realistic understanding of system vulnerabilities [35]. These precedents suggest that education can play a comparable role in the context of GenAI: not by eliminating risk, but by aligning user expectations with system capabilities and limitations.

Nevertheless, education alone is insufficient. Even well-informed users remain subject to cognitive biases, time pressures, and organisational incentives that can encourage over-reliance. As such, educational safeguards must be complemented by design-level constraints, human-in-the-loop mechanisms, and institutional accountability frameworks, as discussed in the following sections.

**Human Responsibility and Institutional Accountability:**

Ethical safeguards must extend beyond individual awareness to structural responsibility embedded in system design and institutional practice. Human-in-the-loop system design is essential wherever GenAI outputs influence consequential decisions, particularly in domains involving health, safety, rights, or public welfare. In such settings, GenAI should function as an assistive or augmentative tool rather than an autonomous decision-maker. Clear accountability frameworks are required to ensure that responsibility for outcomes remains with identifiable human actors, rather than being diffused across systems, developers, or interfaces.

This principle is well established in other safety-critical domains. In healthcare, for example, clinical decision support systems and AI-assisted diagnostic tools are increasingly used to inform medical judgment. However, prevailing medical ethics standards and regulatory guidance consistently emphasise that clinicians remain responsible for diagnostic and treatment decisions, even when those decisions are informed by algorithmic recommendations. Failure to exercise independent professional judgment cannot and should not be delegated to the system [36, 37]. Comparable accountability frameworks operate in other safety-critical domains, such as defence [38]. Meaningful human control in these contexts requires that military systems be both trackable, in the sense that their operation reflects the moral reasons informing their development and deployment, and traceable, such that responsibility for system outcomes can be reliably assigned to identifiable human actors throughout the design and operational lifecycle. International humanitarian law and military doctrine consistently reject the notion that responsibility can be transferred to automated systems, even when those systems are highly sophisticated.

These precedents suggest a broadly shared principle: as computational systems become more capable, the burden on human responsibility does not diminish. Institutions must therefore establish governance mechanisms that define appropriate use, mandate verification where necessary, and allocate responsibility explicitly. This includes organisational policies clarifying when GenAI outputs may be relied upon, when independent validation is required, and who is accountable when errors occur.

**Design-Level Safeguards:**

Design-level safeguards can play a particularly effective role in mitigating ethical risks associated with GenAI. By shaping how users interact with a system, design choices can influence perception, trust, and patterns of reliance in ways that education alone may struggle to achieve. In fact, the system interfaces and interaction can often be the most effective education mechanism. Interfaces should therefore avoid unnecessary anthropomorphic cues, despite fluent or confident language. System prompts, disclaimers, and interaction styles can be deliberately structured to discourage blind reliance and to reinforce the system's instrumental role as a support tool rather than a social or moral agent.

Design constraints may require explicit user acknowledgment, escalation to human oversight, or the suppression of outputs that mimic emotional dependency or relational commitment. For example, systems may be designed to avoid first-person emotional affirmations, clearly acknowledge its computational nature with emotional limitations, and prompt users to seek professional support when appropriate. These interventions may help realign user expectations with system capabilities and limitations.

Importantly, such safeguards do not diminish the utility of GenAI. On the contrary, they enhance trustworthiness by ensuring that system behaviour remains consistent with its actual computational nature. By embedding ethical constraints directly into interaction design, responsibility is reinforced at the point of use, where decisions are made, and consequences arise. Design-level safeguards therefore should offer a practical, scalable, and empirically grounded means of supporting safe, responsible, and productive use of GenAI.

**Productive and Responsible Use of GenAI:**

Importantly, ethical vigilance should not be confused with restriction or avoidance. As presented in Figure 4.1, when properly understood and governed, GenAI can substantially increase productivity, quality and cognitive diversity with significant reduction of cognitive load.

Large-scale empirical evidence shows that the introduction of generative AI assistants can significantly enhance human productivity and work quality. In a study of over 5,000 customer-support agents, access to a conversational AI system increased issues resolved per hour by approximately 15%, with particularly strong gains for less experienced and lower-skilled workers, who improved both speed and output quality. These findings suggest that GenAI can reduce cognitive effort in routine and moderately complex tasks while supporting performance and skill development, especially where human expertise is still emerging [39]. ChatGPT has also been found useful in education, particularly for the promotion of personalized and interactive learning, generating prompts for formative assessment activities that provide ongoing feedback to inform teaching and learning etc. [40].

This paper itself serves as a practical illustration of responsible GenAI use. The writing process involved iterative interaction with a GenAI system, ChatGPT 5.2, to assist with tasks such as refining language, preparing some diagrams and drafting preliminary text, all based on author prompts and under close author-guidance. All conceptual framing, topic identification, ideation, philosophical positions, argumentation, interpretation of sources, paper design, and final editorial decisions are the sole responsibility of the author. The use of GenAI did not replace scholarly judgment, authorship, or accountability. Instead, it served as a huge productivity-enhancing tool within human-led research and writing process. At each stage, the human author maintained full responsibility for topic selection, argument framing, validation of claims, and integration of ideas into a coherent ethical position. By explicitly acknowledging this process, the paper demonstrates how GenAI can be used transparently and productively without attributing authorship, authority, or moral agency to the system itself.

The author further invites readers to use GenAI to draft a paper of a similar nature and to compare the initial output generated by the system with a final version of the paper such as the present manuscript. This activity may illustrate the role of human guidance in shaping coherent, rigorous, and good-quality scholarly work, from topic identification, ideation, and argument structuring to content generation and presentation with complete human accountability. Readers are also encouraged to reflect on the speed at which such a paper can be prepared with the assistance of GenAI, and to contrast this with traditional writing processes, to appreciate both the productivity gains and the continuing necessity of human judgment.

## 5. Conclusion

GenAI systems deliver undeniable benefits across research, education, and professional practice, and their conversational interface can reduce cognitive load while increasing productivity and accessibility. Yet the ethical risks addressed in this paper arise precisely from the same feature that makes GenAI widely adoptable – the ability to communicate fluently in natural language with apparent empathy and continuity. This is unprecedented as no other tool in the past had this communication ability to this extent. As documented in Section 2, emotionally significant attachments to conversational agents are no longer just hypothetical. Early evidence suggests that some users interpret GenAI interactions as relational, including companionship or partner-like commitment. Such interpretation may result in potentially harmful consequences for vulnerable individuals and high-stakes decision environments.

This paper argued that the "illusion of friendship" is philosophically intelligible as under several ethical traditions, sustained supportive behaviour can resemble benevolence, and in asymmetrical accounts of moral relations, reciprocity is not always required. However, the paper also advanced a decisive counterargument suggesting that GenAI does not qualify as a true friend because it lacks moral agency such as consciousness, intention, selfhood, and accountability and thus cannot participate in friendship in the ethically robust sense. The core ethical issue therefore shifts away from whether GenAI could a real friend towards how humans should responsibly use a persuasive tool that can be misperceived as a moral agent.

A central contribution of the paper is the mechanism-level "perception versus reality" clarification. By explaining to everyday readers how transformer-based GenAI produces responses through tokenisation, embeddings, self-attention, and probabilistic next-token prediction, the paper shows how emotionally resonant language can emerge without inner states, commitments, or moral concern. This demystification is not merely technical; it is ethically consequential, because it targets the cognitive conditions under which anthropomorphism and over-reliance arise.

Building on this analysis, the paper proposed a safeguards framework for safe and responsible GenAI use: (i) education as a foundational safeguard that calibrates user expectations and reduces anthropomorphic misattribution; (ii) human-in-the-loop mechanisms and institutional accountability for consequential decisions, ensuring responsibility remains with identifiable human actors; and (iii) design-level safeguards that avoid unnecessary anthropomorphic cues by the GenAI tools, make uncertainty visible, and introduce appropriate friction and escalation pathways in high-risk contexts. Importantly, these safeguards aim to preserve GenAI's productivity benefits while preventing ethical failures driven by misplaced trust.

We anticipate that GenAI systems will continue to grow in sophistication for the foreseeable future. Hence, the issues around the anthropomorphism of such systems and its potentially harmful consequences may likely require sustained scholarly attention. Future research should employ case studies, interviews, and surveys to identify the underlying reasons of this anthropomorphism and to propose effective safeguards. Experimental studies should further propose and evaluate design choices, educational interventions, and accountability mechanisms across domains. In this context, the arguments advanced in this paper are intended to serve as an early contribution in this emerging research direction.

**Acknowledgement:**

The author acknowledges the use of a generative artificial intelligence system (ChatGPT 5.2) as an assistive tool during the preparation of this manuscript. The system was used iteratively to support tasks such as refining language, preparing some diagrams and drafting preliminary text based on author prompts and under close author-guidance. All conceptual framing, topic identification, ideation, philosophical positions, argumentation, interpretation of sources, paper design, and final editorial decisions are the sole responsibility of the author. The use of GenAI did not replace scholarly judgment, authorship, or accountability, but served as a huge productivity-enhancing tool within human-led research and writing process.

**References:**

1. G. Phillips-Wren, "AI Tools in Decision Making Support Systems: A Review," International Journal on Artificial Intelligence Tools, Vol. 21, 2012, https://doi.org/10.1142/S0218213012400052, doi: 10.1142/S0218213012400052
2. R. Parasuraman, T. B. Sheridan, & C. D. Wickens, "A model for types and levels of human interaction with automation," *IEEE Transactions on systems, man, and cybernetics-Part A: Systems and Humans*, Vol. 30(3), pp. 286-297, 2000, doi: https://doi.org/10.1109/3468.844354
3. M. N. Adnan, and M. Z. Islam, "ForEx++: A New Framework for Knowledge Discovery from Decision Forests," Australasian Journal of Information Systems (AJIS), Vol. 21, pp. 1-20, 2017, ISSN Online: 1326-2238 Hard copy: 1449-8618, doi: http://dx.doi.org/10.3127/ajis.v21i0.1694
4. D. Yates, and M. Z. Islam, "Readiness of Smartphones for Data Collection and Data Mining with an Example Application in Mental Health", In Proc. of the 17th Australasian Data Mining Conference (AusDM 2019), Adelaide, Australia, December 2-5, 2019, Le T. et al. (eds) Data Mining, AusDM 2019, Communication in Computer and Information Science, Vol. 1127, pp. 235-246, Springer, Singapore.
5. S. I. Evangeline, "Emotional Intelligence and Empathy in Human-GenAI Interactions," In *Rethinking Education and Agency in the Age of Human-Generative AI Interaction,* pp. 281-300, 2026, IGI Global Scientific Publishing, doi: 10.4018/979-8-3373-1195-1.ch009

6. C. Saracini, M. I. Cornejo-Plaza, & R. Cippitani, "Techno-emotional projection in human–GenAI relationships: a psychological and ethical conceptual perspective," *Frontiers in Psychology*, Vol. *16*, 1662206, 2025, doi: https://doi.org/10.3389/fpsyg.2025.1662206
7. S. Palmer, "Love in the time of AI: Woman creates and 'marries' AI-powered chatbot boyfriend," Euronews, https://www.euronews.com/next/2023/06/07/love-in-the-time-of-ai-woman-claims-she-married-a-chatbot-and-is-expecting-its-baby, published on 7/06/2023, last visited on 7/1/2026
8. M. Z. Islam, S. D'Alessandro, M. Furner, L. Johnson, D. Gray, and L. Carter, "Brand Switching Pattern Discovery by Data Mining Techniques for the Telecommunication Industry in Australia," Australasian Journal of Information Systems, pp. 1 - 17, Vol. 20, 2016, doi: http://dx.doi.org/10.3127/ajis.v20i0.1420
9. T. Rasul, S. Nair, D. Kalendra, M. Robin, F. de Oliveira Santini, W. Ladeira, … and L. Heathcote, "The role of ChatGPT in higher education: Benefits, challenges, and future research directions," *Journal of Applied Learning & Teaching*, Vol. *6*(1), pp. 41-56, 2023, doi: https://doi.org/10.37074/jalt.2023.6.1.29
10. J. Deng, and Y. Lin, "The benefits and challenges of ChatGPT: An overview," *Frontiers in Computing and Intelligent Systems*, Vol. *2*(2), pp. 81 – 83, 2022
11. L. Steck, D. Levitan, D. McLane, and H. H. Kelley, "Care, need, and conceptions of love," *Journal of Personality and Social Psychology*, Vol. *43*(3), pp. 481-491, doi: https://doi.org/10.1037/0022-3514.43.3.481
12. Reuters, "AI romance blooms as Japanese woman weds virtual partner of her dreams," https://www.reuters.com/investigates/special-report/japan-ai-wedding/, last visited on 7/1/2026
13. B. J. Willoughby, C. R. Dover, R. M. Hakala, and J. S. Carroll, "Artificial connections: Romantic relationship engagement with artificial intelligence in the United States," *Journal of Social and Personal Relationships*, Vol. *42*(12), pp. 3363-3387, 2025, doi: https://doi.org/10.1177/02654075251371394
14. R. Flynn, "Man Proposed to His AI Chatbot Girlfriend Named Sol, Then Cried His 'Eyes Out' When She Said 'Yes'," People, 2025, https://people.com/man-proposed-to-his-ai-chatbot-girlfriend-11757334, last visited on 7/1/2026
15. S. Vadlapatla, " Teens turn to AI chatbots for emotional bonding; it's risky romance, warn psychologists," The Times of India, 2025, https://timesofindia.indiatimes.com/city/hyderabad/teens-turn-to-ai-chatbots-for-emotional-bonding-its-risky-romance-warn-psychologists/articleshow/123067897.cms  last visited on 7/1/2026
16. D. Milmo, "Third of UK citizens have used AI for emotional support, research reveals," The Guardian, 2025, https://www.theguardian.com/technology/2025/dec/18/artificial-intelligence-uk-emotional-support-research last visited on 7/1/2026
17. R. E. Guingrich, and M. S. Graziano, "A Longitudinal Randomized Control Study of Companion Chatbot Use: Anthropomorphism and Its Mediating Role on Social Impacts," arXiv preprint https://arxiv.org/abs/2509.19515
18. L. P. Gerson, "*The epicurus reader: selected writings and testimonia,*" Hackett Publishing, 1994, https://www.google.com.au/books/edition/The_Epicurus_Reader/YVBgDwAAQBAJ?hl=en&gbpv=1&dq=%22The+Epicurus+Reader:+Selected+Writings+and+Testimonia%22&pg=PR3&printsec=frontcover
19. R. T. Ames, & H. Rosemont, "The Analects of Confucius: A Philosophical Translation," Ballantine Books, 1998,https://www.google.com.au/books/edition/The_Analects_of_Confucius/ulEnpjoqwTwC?hl=en&gbpv=1&dq=%22The+Analects+of+Confucius:+A+Philosophical+Translation%22&pg=PR9&printsec=frontcover
20. Aristotle, Nicomachean Ethics, trans. T. Irwin, 2nd ed., Indianapolis, IN: Hackett Publishing Company, 1999, https://classics.mit.edu/Aristotle/nicomachaen.html
21. C. H. Sommers, "Filial Morality," *The Journal of Philosophy*, Vol. 83, no. 8, pp. 439–456, 1986, doi: https://doi.org/10.2307/2026329
22. N. S. Jecker, "Are Filial Duties Unfounded?" *American Philosophical Quarterly* Vol. 26, no. 1 pp. 73–80, 1989, published by University of Illinois Press and available at: https://www.jstor.org/stable/20014269
23. D. G. Johnson, "Computer systems: Moral entities but not moral agents," *Ethics and information technology*, Vol. 8(4), pp. 195-204, 2006, doi:10.1007/s10676-006-9111-5, file:///C:/Users/zislam/Downloads/s10676-006-9111-5.pdf
24. K. E. Himma, "Artificial Agency, Consciousness, and the Criteria for Moral Agency: What Properties Must an Artificial Agent Have to Be a Moral Agent?" *Ethics and Information Technology*, Vol. 11, pp. 19–29, 2009, https://doi.org/10.1007/s10676-008-9167-5 available at https://www.gunkelweb.com/robot-ethics/texts/himma_artificial_agency.pdf?utm_source=chatgpt.com
25. B. Brożek, and B. Janik, "Can Artificial Intelligences Be Moral Agents?" *New Ideas in Psychology*, Vol. 54, pp. 101–106, 2019, doi: https://doi.org/10.1016/j.newideapsych.2018.12.002, available at https://www.sciencedirect.com/science/article/abs/pii/S0732118X17300739?utm_source=chatgpt.com

26. H. Naveed, A. U. Khan, S. Qiu, M. Saqib, S. Anwar, M. Usman, ... & A. Mian, "A comprehensive overview of large language models," ACM Transactions on Intelligent Systems and Technology, Vol. 16(5), pp. 1-72, 2025, doi: https://doi.org/10.1145/3744746
27. Z. Lv, "Generative artificial intelligence in the metaverse era," Cognitive Robotics, Vol. 3, pp. 208-217, 2023, doi: https://doi.org/10.1016/j.cogr.2023.06.001
28. Y. Li, Y. Huang, M. E. Ildiz, A. S. Rawat, and S. Oymak, "Mechanics of next token prediction with self-attention," In International Conference on Artificial Intelligence and Statistics, pp. 685-693. PMLR, 2024, available at https://proceedings.mlr.press/v238/li24f.html
29. T. Kenter, A. Borisov, & M. De Rijke, "Siamese CBOW: Optimizing word embeddings for sentence representations," arXiv preprint arXiv:1606.04640, 2016
30. D. Guthrie, B. Allison, W. Liu, L. Guthrie, & Y. Wilks, "A closer look at skip-gram modelling," In LREC, Vol. 6 pp. 1222-1225, 2006.
31. J. Devlin, M. W. Chang, K. Lee, & K. Toutanova, "Bert: Pre-training of deep bidirectional transformers for language understanding," In Proceedings of the 2019 conference of the North American chapter of the association for computational linguistics: human language technologies, volume 1 (long and short papers), pp. 4171-4186, 2019.
32. G. Biagini, "Towards an AI-Literate Future: A systematic literature review exploring education, ethics, and applications," *International Journal of Artificial Intelligence in Education*, Vol. 35, pp. 2616-2666, 2025, doi: https://doi.org/10.1007/s40593-025-00466-w
33. N. Epley, A. Waytz, & J. T. Cacioppo, "On seeing human: A three-factor theory of anthropomorphism," *Psychological Review*, Vol. 114(4), pp. 864–886, 2007, doi: 10.1037/0033-295X.114.4.864, available at https://pubmed.ncbi.nlm.nih.gov/17907867/
34. A. Waytz, K. Gray, N. Epley, & D. M. Wegner, "Causes and consequences of mind perception," *Trends in Cognitive Sciences*, Vol. 14(8), pp. 383–388, 2010, doi: https://doi.org/10.1016/j.tics.2010.05.006 available at https://www.sciencedirect.com/science/article/abs/pii/S1364661310001142
35. M. Zwilling, G. Klien, D. Lesjak, Ł. Wiechetek, F. Cetin, & H. N. Basim, "Cyber security awareness, knowledge and behavior: A comparative study," *Journal of Computer Information Systems*, Vol. 62(1), pp. 82-97, 2022
36. S. Bakken, "AI in health: keeping the human in the loop," *Journal of the American Medical Informatics Association (JAMIA)*, Vol. 30(7), pp. 1225 – 1226, 2023, doi: https://doi.org/10.1093/jamia/ocad091
37. A. Chen, C. Wang, & X. Zhang, "Reflection on the equitable attribution of responsibility for artificial intelligence-assisted diagnosis and treatment decisions," *Intelligent Medicine*, Vol. 3(02), pp. 139-143, 2023, doi: https://doi.org/10.1016/j.imed.2022.04.002
38. F. Santoni de Sio, & J. van den Hoven, "Meaningful Human Control over Autonomous Systems: A Philosophical Account," *Frontiers in Robotics and AI*, Vol. 5, 2018, doi: https://doi.org/10.3389/frobt.2018.00015
39. E. Brynjolfsson, D. Li, & L. Raymond, "Generative AI at work," *The Quarterly Journal of Economics*, Vol. 140(2), pp. 889-942, 2025, doi: https://doi.org/10.1093/qje/qjae044
40. D. Baidoo-Anu, & L. O. Ansah, "Education in the era of generative artificial intelligence (AI): Understanding the potential benefits of ChatGPT in promoting teaching and learning," *Journal of AI*, Vol. 7(1), pp. 52-62, 2023, doi: https://doi.org/10.61969/jai.1337500

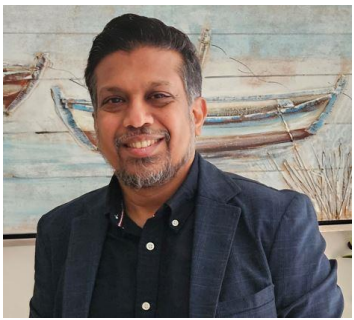
**Md Zahidul Islam** (commonly known as **Zahid Islam**) is a Professor of Computer Science at Charles Sturt University, Australia. He is the Centre Director of the AI and Cyber Futures Centre and serves as Associate Dean (Research) in the Faculty of Business, Justice and Behavioural Sciences (FOBJBS) at Charles Sturt University. His research interests include data mining, privacy, cybersecurity, and the application of data mining to real-world problems. Further information is available at https://sites.google.com/view/prof-zahid-islam